
\documentclass[onecolumn]{aastex61}
\usepackage{graphicx}
\usepackage{soul}

\usepackage{CJK}

\revised{\today}
\shorttitle{Solar Cycle Scale Trend in Meridional Flow}
\shortauthors{Mahajan, Sun \& Zhao}

\begin{document}
\begin{CJK*}{UTF8}{gbsn}

\title{Removal Of Active Region Inflows Reveals a Weak Solar Cycle Scale Trend In Near-surface Meridional Flow}

%% OR: TDH Flows around active regions: I. Determination of background near-surface flows
%% TDH Flows around active regions: II. Dependence of inflows on properties of ARs

\correspondingauthor{Sushant S. Mahajan}
\email{mahajans@stanford.edu}

\author[0000-0003-1753-8002]{Sushant S. Mahajan}
\affil{Institute for Astronomy, University of Hawaii, Manoa, HI, USA}
\affil{W.W. Hansen Experimental Physics Laboratory, Stanford University, Stanford, CA 94305-4085, USA}

\author[0000-0003-4043-616X]{Xudong Sun (孙旭东)}
\affil{Institute for Astronomy, University of Hawaii, Manoa, HI, USA}

\author[0000-0002-6308-872X]{Junwei Zhao (赵俊伟)}
\affil{W.W. Hansen Experimental Physics Laboratory, Stanford University, Stanford, CA 94305-4085, USA}

%--------------------------------------------------------------------------------
%--------------------------------ABSTRACT----------------------------------------
%--------------------------------------------------------------------------------

\begin{abstract}
Using time-distance local helioseismology flow maps within 1 Mm of the solar photosphere, we detect inflows toward activity belts that contribute to solar cycle scale variations in near-surface meridional flow. These inflows stretch out as far as 30 degrees away from active region centroids. If active region neighborhoods are excluded, the solar cycle scale variation in background meridional flow diminishes to below 2~m~s$^{-1}$, but still shows systematic variations in the absence of active regions between Sunspot Cycles 24 and 25. We, therefore, propose that the near-surface meridional flow is a three component flow made up of: a constant baseline flow profile that can be derived from quiet Sun regions, variations due to inflows around active regions, and solar cycle scale variation of the order of 2~m~s$^{-1}$. Torsional oscillation, on the other hand, is found to be a global phenomenon i.e. exclusion of active region neighborhoods does not affect its magnitude or phase significantly. This non-variation of torsional oscillation with distance away from active regions and the three-component breakdown of the near-surface meridional flow serve as vital constraints for solar dynamo models and surface flux transport simulations.
\end{abstract}


%--------------------------------------------------------------------------------
%--------------------------------INTRODUCTION------------------------------------
%-------------------------------------------------------------------------------

\section{Introduction} \label{sec:intro}

The Sun's global flows have now been studied for more than three decades. The solar rotation rate is known to vary in latitude and depth, thus justifying the use of the term ``differential rotation'' \citep{1858MNRAS..19....1C}. The rotation rate of the Sun, however, isn't constant in time. Its temporally varying component is called torsional oscillation \citep{1980ApJ...239L..33H,1994SoPh..149..417K,1998ApJ...505..390S,2000SoPh..192..427H,2000SoPh..192..437T,2004ESASP.559..468H}.
Many studies have measured torsional oscillation on or beneath the surface of the Sun over different periods of time \citep{1980ApJ...239L..33H,1982SoPh...75..161L,2002Sci...296..101V,2004ApJ...603..776Z,2009ApJ...701L..87H,2011ApJ...729...80H,2021ApJ...917..100M}. It appears as a pattern of slower and faster than average rotation bands propagating toward the equator at low latitudes and toward the poles at high latitudes. This torsional oscillation pattern is speculated to be linked to the extended solar cycle \citep{1988Natur.333..748W,2014ApJ...792...12M,2017FrASS...4....4M}. However, to measure the torsional oscillation accurately, we do not yet know what the correct choice of background rotation rate is, as the temporal mean of torsional oscillation is not %consistent 
constant from one solar cycle to another (\citet{2013ApJ...767L..20H}). In fact, choosing the background rotation rate as an average over Solar Cycle 23 led to the apparent absence of the high-latitude branch preceding Cycle 25 \citep{2018ApJ...862L...5H}, which can be seen if one chooses a more carefully constructed background rotation rate. 

The Sun also exhibits a North-South flow on the surface, that is much weaker ($\approx$~15~m~s$^{-1}$) than its rotation rate and transports magnetic flux away from the equator in both hemispheres. This flow is known as meridional flow \citep{1979SoPh...63....3D,1982SoPh...80..361L,1988SoPh..117..291U}. Many studies have reported that the meridional flow, like torsional oscillation, varies throughout the solar cycle too  \citep{1993SoPh..147..207K,1998ApJ...508L.105B,2003PhDT.........9G,2004ApJ...603..776Z,2008SoPh..252..235G,2010Sci...327.1350H,2011ApJ...729...80H,2021ApJ...917..100M}. 
However, it has been difficult to identify and isolate the time-varying component of both these flows. It is unclear whether a relatively constant background meridional flow (if it exists), can be approximated by the temporal average, or with the flow profile at solar minima/maxima.

Over the past two decades, regular observations available from Michelson Doppler Imager \citep[MDI;][]{1995SoPh..162..129S} onboard the Solar Heliospheric Observatory, the Global Oscillations Network Group \citep[GONG;][]{1996Sci...272.1284H} and the Helioseismic Magnetic Imager \citep[HMI;][]{2012SoPh..275..207S} instrument onboard the {\it Solar Dynamics Observatory} led to the speculation that the variation in meridional flow can be explained by inflows toward active region centers \citep{2004SoPh..224..217G,2010ApJ...720.1030C}. However, this was not confirmed by \citet{2008SoPh..252..235G} and \citet{2020SoPh..295...47K} where they found that the solar-cycle-scale variations in meridional flow do attenuate by masking areas around active regions, but significant variations remain in quiet-Sun regions. The reported strength (from 10 to 50~m~s$^{-1}$) and the reported spatial extent (10 to 30 degrees heliographic) of these inflows toward active regions can vary depending on the measurement technique, the resolution of velocity inversions, the active region detection criterion as well as the size of area around active regions that was masked. A deeper study into understanding how the reported inflows could vary based on one or all of these factors is therefore necessary.

On the other hand, the torsional oscillation pattern is also generally considered to be linked to inflows into active regions. However, to get a torsional oscillation type behavior from inflows, these inflows would have to be asymmetric in longitude (East-West) around the centers of active regions. There is some indication that this might be the case in measurements of flows around active regions derived through helioseismic holography \citep{2019ApJ...873...94B}.

With the advent of local helioseismology \citep{1988ApJ...333..996H,1993Natur.362..430D} and the high resolution velocity inversions derived from HMI data since 2010, we now have one full solar cycle of velocity maps (2010--2021) with unprecedented resolution to help us answer some of these questions \citep{2012SoPh..275..375Z}. In this paper, we present an analysis of the strength and nature of meridional flow and torsional oscillation as a function of distance away from active region flux centroids. We are able to infer that 1) near-surface torsional oscillation is not spatio-temporally linked to the presence of active regions as the East-West inflows into active regions cancel out; 2) the near-surface meridional flow can be modeled as a three component flow: a constant baseline meridional flow that can be derived by measuring the North-South flow at least 30 degrees and 48 hours away from all active regions, inflows into active regions, and a weak ($\sim 2$ m/s) solar-cycle-scale variation in quiet-Sun regions that appears like an equatorward propagating band in each hemisphere.

We describe the data we used in Sec.\ \ref{sec:data}, our methodology in Sec.\ \ref{sec:methodology} and our results in Sec.\ \ref{sec:results}. A more detailed description of corrections for systematic errors in the data is given in Appendix \ref{sec:systematics}. In Sec.\ \ref{sec:discussion} we discuss the potential benefits, use cases of our results, and how they add to the current physical understanding of our closest star. 

\section{Data} \label{sec:data}

\subsection{Time-Distance Helioseismic Velocity Maps}

The HMI's time--distance helioseismic data-analysis pipeline, developed soon after the launch of {\it SDO}, produces the Sun's near-full-disk subsurface flow maps every 8 hours \citep{2012SoPh..275..375Z}. The flow maps, spanning approximately $123\degr \times 123\degr$ in both longitude and latitude, cover horizontal flow fields from the photosphere to about 20\,Mm in depth. The shallowest results from this pipeline, with a depth of 0 -- 1 Mm, are used in this analysis. For the time--distance helioseismic analysis, the full-disk observations are divided into $5 \times 5$ patches, each patch covering about a $30\degr \times 30\degr$ area with some spatial overlap in between as shown in Fig.\ 3 of \citet{2012SoPh..275..375Z}. The measurements and inversions are carried out individually in each patch, and the 25 sets of final results of flows are then merged together to form the full-disk flow maps. Typically, these subsurface flow maps can well resolve supergranular flows, and are often averaged in longitude to derive the latitude-dependent rotation and meridional-flow profiles in the Sun's shallow interior \citep[e.g.,][]{2014ApJ...789L...7Z}. The spatial sampling for these flow maps is $0\fdg12$~pixel$^{-1}$, but the actual spatial resolution is believed a few times larger than that due to the long wavelengths of the $p$-mode waves used in the measurements. For uncertainties about these  subsurface flow maps, please refer to \citet{2012SoPh..275..375Z}.

The full-disk subsurface flow maps ($V_{x}, V{y}$) and the co-aligned magnetograms, which  match the flow maps both spatially and temporally, are available since May 2010 to the present (we use them until Aug 2022). The total number of velocity maps available are 13431, spaced evenly every 8 hours with certain gaps during calibration runs and %Moon XS: I think it's mostly the Earth that blocks the Sun rather than the Moon.
eclipses. These flow maps carry some systematic errors, such as the systematic center-to-limb effect. Explanation of the corrections of systematic errors in data are described in the appendix.

The typical uncertainties in the determination of velocities are given in Table 3 of \citet{2012SoPh..275..375Z} viz. $7.8$ m~s$^{-1}$ for quiet Sun regions and $58.3$ m~s$^{-1}$ in active regions. While these seem fairly large compared to the typical amplitude of meridional flow and torsional oscillation, they become minuscule (less than $0.1$ m~s$^{-1}$) when the data is averaged in longitude and binned every Carrington rotation.

%Velocity maps ($V_{x}, V{y}$) coalligned with magnetograms are available in Postel projection ($-61.5\degr$ to $+61.5\degr$ in latitude and longitude) from the HMI time-distance helioseismology pipleline \citep{2012SoPh..275..375Z} at a cadence of 8 hours since May 2010 to present.
%Explanation of the corrections of systematic errors in data are described in the appendix.

%The total number of velocity maps available are 11,910 spaced every eight hours with certain gaps during calibration runs and Moon eclipses.

\subsection{SHARP Metadata}

In order to identify active regions (ARs) in the velocity maps, we query active region metadata from the
Space-Weather HMI Active Region Patches \citep[SHARPs;][]{2014SoPh..289.3549B}. The data series (available from JSOC: \url{jsoc.stanford.edu}) comprises vector magnetic field maps along with metadata describing the properties of these active regions like their location, area, net unsigned flux, etc. For this study, we use the magnetic flux weighted centroid latitude and longitude (LAT\_FWT and LON\_FWT) as location coordinates.

Each SHARP region reported by the HMI data pipeline is given a unique identifying number called HARPNUM. Even though data for SHARP regions ranging from 1 to 8578 in HARPNUM are available until August 3, 2022 00:00:00 TAI, 3657 out of these are empty, i.e., have no metadata. This is because these were created as part of the near-real-time (NRT) series and were later found to be in close proximity to other active regions, so they were merged into a larger composite SHARP region when the definitive series was created. Out of the remaining 4921 SHARP regions: 1357 always have one matching NOAA AR from the RGO/NOAA database throughout their lifetimes. 267 SHARP regions match multiple NOAA ARs whereas 25 SHARP regions match a varying number of NOAA ARs depending on the timestamp and NOAA AR reports (may be missing some AR in the RGO/NOAA catalogue). The remaining 3272 (66.5\%) of definitive series SHARP regions have no matching NOAA AR. This indicates that the automated SHARP region detection routine is better suited for the identification of ARS for our goals as it detects much smaller concentrations of magnetic flux than the RGO/NOAA catalogue. This helps us be very strict at the exclusion of active regions in the calculation of the background meridional and zonal flows.

We make the following three corrections to the SHARP metadata. First, around 3.1\% of LAT\_FWT and LON\_FWT values are missing for the first few records of some active regions, when the regions were either too small or yet to emerge. We %replace
specify the locations in such records by keeping the LAT\_FWT constant and extrapolating LON\_FWT using a standard solar differential rotation profile \citep{1984SoPh...90..199S}. Second, in cases where multiple NOAA designated active regions (NOAA\_ARS) were reported inside a SHARP region, we query their individual locations from the RGO/NOAA sunspot database \citep{RGO} maintained by David H. Hathaway at \url{www.solarcyclescience.com} and treat them as separate active regions. This helps properly center the inflows into active regions around their individual centroids if multiple regions are present close to each other.  Third, we extrapolate the locations of all active regions to include up to 48 hours before their first record and 48 hours after their last record. This is done to avoid the potential contamination of background flow measurements due to changing flow patterns prior to the emergence of active regions as well as residual flow patterns after their disappearance.

\section{Methodology} \label{sec:methodology}

\begin{figure*}[!ht]
    \centering
    \includegraphics[width = 10 cm]{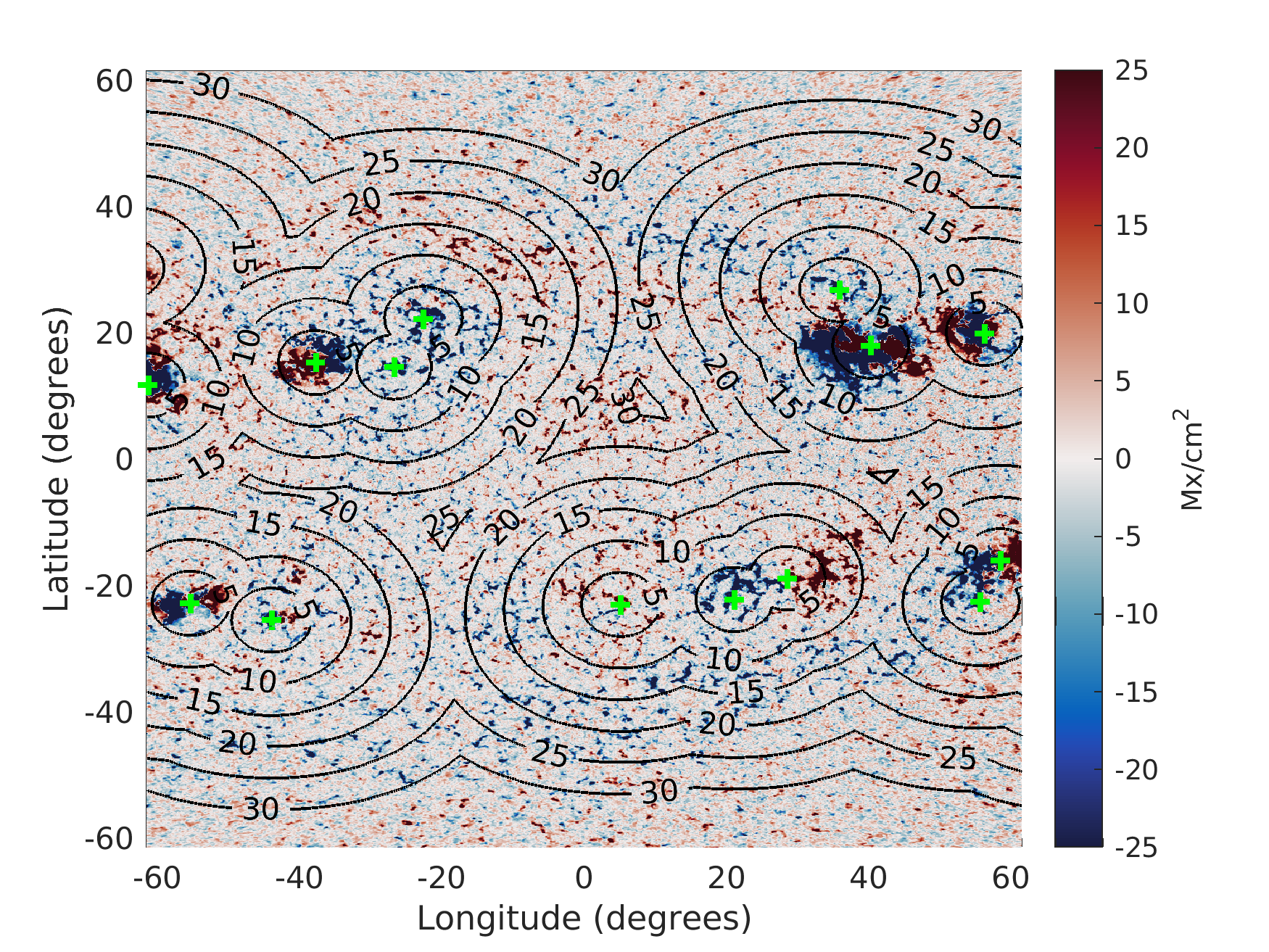}
    \caption{A magnetogram (magnetic field values represented by the color bar) from 2012 March 12 04:00:00 TAI with annuli around active regions marked by contours in steps of $5\degr$ away from active region centroids (marked by green plus).}
    \label{fig:annuli_composite}
\end{figure*}

%In order to analyze flows around active regions on a ball of superheated plasma that rotates differentially, the choice of a reference frame is crucial and might have a huge impact on inferences drawn from such an analysis. It is therefore, extremely important to separate spatiotemporal variations in plasma flows over different scales in order to build an unbiased background flow profile that will serve as a comoving reference frame for the analysis of local plasma flows around active regions. In order to separate the background flow from flows perturbed by active regions, we need to first find out the spatial extent to which active regions influence the near-surface flows.

% The way in which flow profiles change with distance away from active regions informs us of the area of influence of active regions as well as the variation in magnitude of perturbation of plasma flows. In some previous studies a preset area ($10$ to $15\degr$) around active regions was analyzed \citep{2016A&A...590A.130L,2019ApJ...873...94B} and therefore, the conclusions might depend on this assumption of the area of influence of active regions because everything outside this preset area was assumed to belong to the background flow that provided the reference frame for the analysis of flows around active regions.

% The contribution of active region inflows to the flows sensitively depends on the analysis technique. To assess the effect of inflows, some previous studies analyzed a preset area ($10$ to $15\degr$) around active regions \citep[e.g.,][]{2016A&A...590A.130L,2019ApJ...873...94B}. 

We choose to analyze the flow profiles in successive annuli around active region centroids with an angular bin width of $5\degr$ up to a distance of $30\degr$ in the heliographic coordinate. A visualization of such annuli around active regions is shown in Fig.\ \ref{fig:annuli_composite}. After masking out all areas outside the annulus that we want to analyze (for e.g., between $5\degr$ to $10\degr$ from active region centroids), we average all velocity maps ($V_{x}$ and $V_{y}$) over each Carrington rotation. Each of these two dimensional (latitude-longitude) velocity maps for every Carrington rotation then goes through systematic corrections described in the Appendix \ref{sec:systematics}. The velocity maps are then averaged over all longitudes to obtain the longitudinally averaged velocity as a function of latitude and time (binned in Carrington rotations). Annual variations due to the orbit of the Earth around the Sun are then removed by fitting sine functions in the temporal dimension at each latitude.  For the annulus range, we stop at the $30\degr$ mark both because we are nearly out of data points, and the influence of active regions on velocities is nearly absent at that distance. Note that most active regions themselves are less than $20\degr$ wide ($\pm 10\degr$). Furthermore, velocity measurements from areas that are over $30\degr$ away from active region centroids are also obtained in the same manner for the calculation of the background flow profiles.

\section{Results}\label{sec:results}

\begin{figure*}
    \flushleft
    \includegraphics[width=\textwidth]{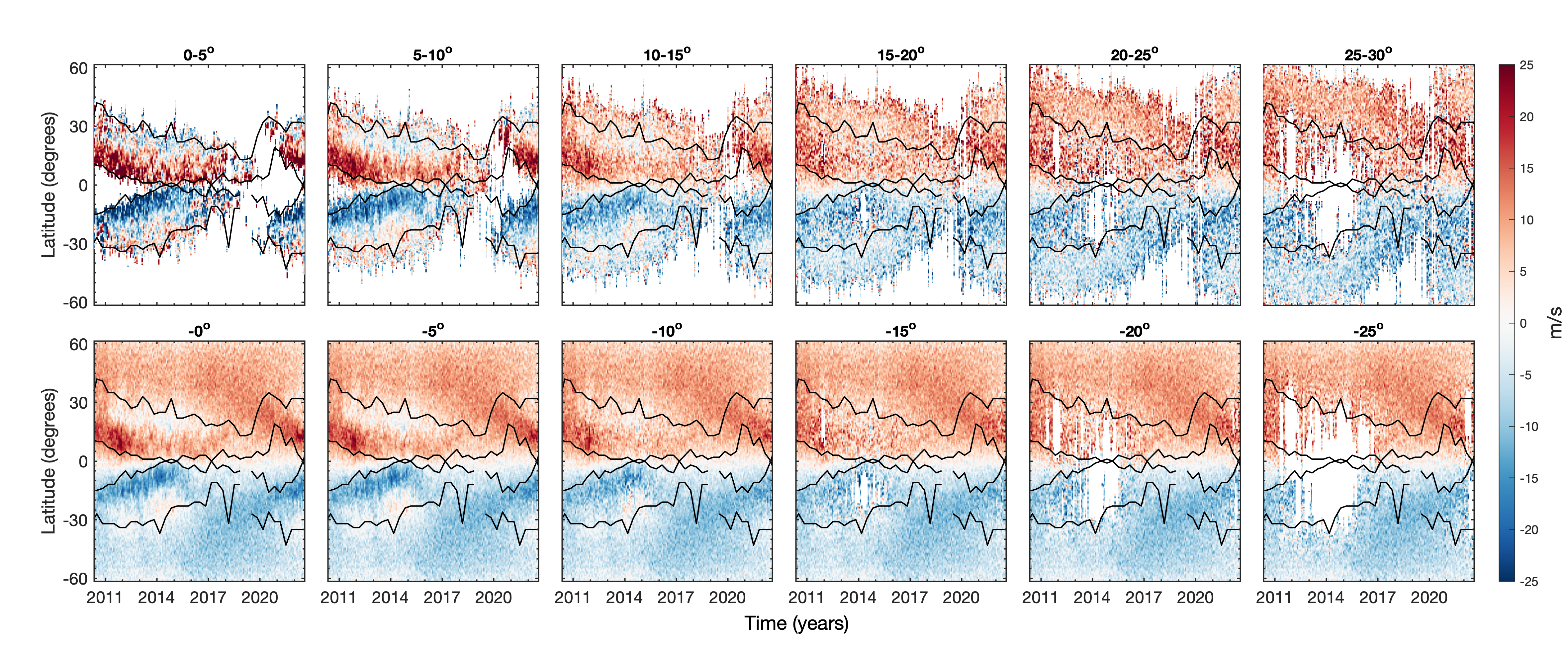}
    \caption{Top row: Meridional flow (MF) derived only from active region neighborhoods, within the annuli marked above each panel, averaged every Carrington rotation (includes regions within 48 hours of the first and last record of active regions). All neighborhood sizes are in heliographic angular distance and measured with respect to the unsigned flux weighted centroid of the active regions reported in SHARP metadata. Bottom row: Background MF derived after removing active region neighborhoods from velocity maps with their sizes marked above (negative sign represents exclusion of these areas). The annuli around active regions show less solar cycle scale variation in MF as the annulus size increases, as does the background MF with larger active region neighborhoods excluded. Here and throughout this manuscript, the black curves denote the active latitudes. Uncertainty decreases (increases) from right to left in top (bottom) row. }
    \label{fig:mf_bfly_annuli}
\end{figure*}

\begin{figure*}
    \centering
    \includegraphics[width=\textwidth]{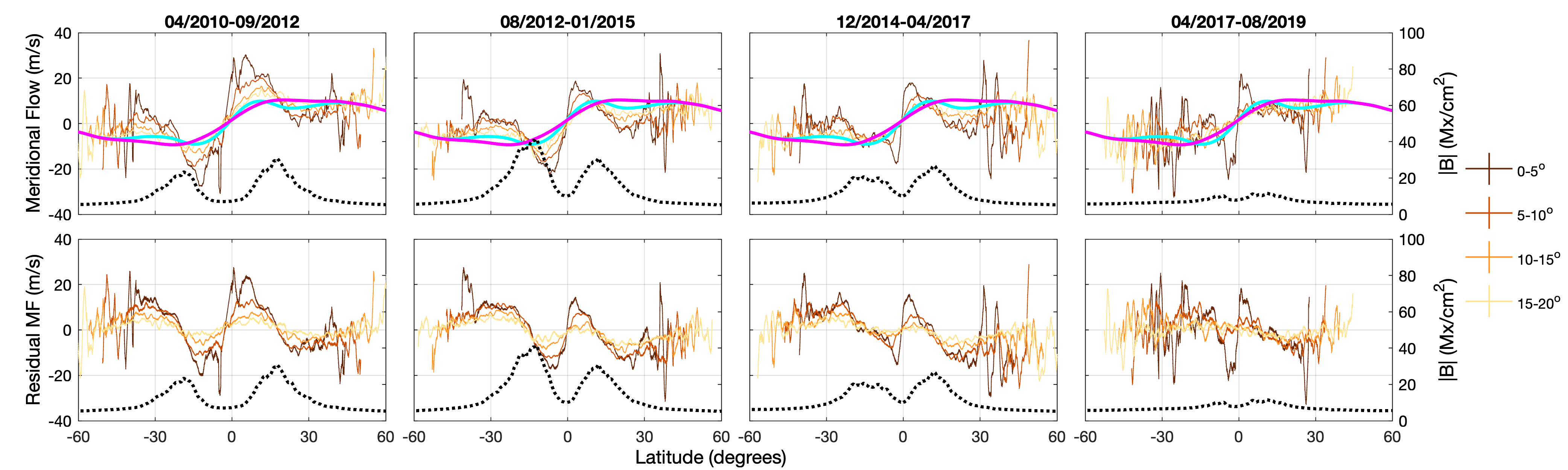}
    \caption{\textit{Top row:} Temporally averaged meridional flow profiles obtained in different annuli around active regions. The baseline meridional flow is shown in magenta whereas the meridional flow averaged over all data is shown in cyan for reference. Meridional flow in different annuli bins are shown in yellow/orange colors. The thickness of the lines depict typical uncertainties ($\sim 1$ m/s closest to active regions and gradually falls to $\sim 0.5$ m/s farthest from active regions). The black dotted line shows the average unsigned magnetic flux density averaged over the same time period in $Mx/cm^{2}$. \textit{Bottom row:} Same as top row, but showing residuals after removal of baseline (magenta) meridional flow profile.}
    \label{fig:mf_curves}
\end{figure*}

\begin{figure}
    \centering
    \includegraphics[width = 10 cm]{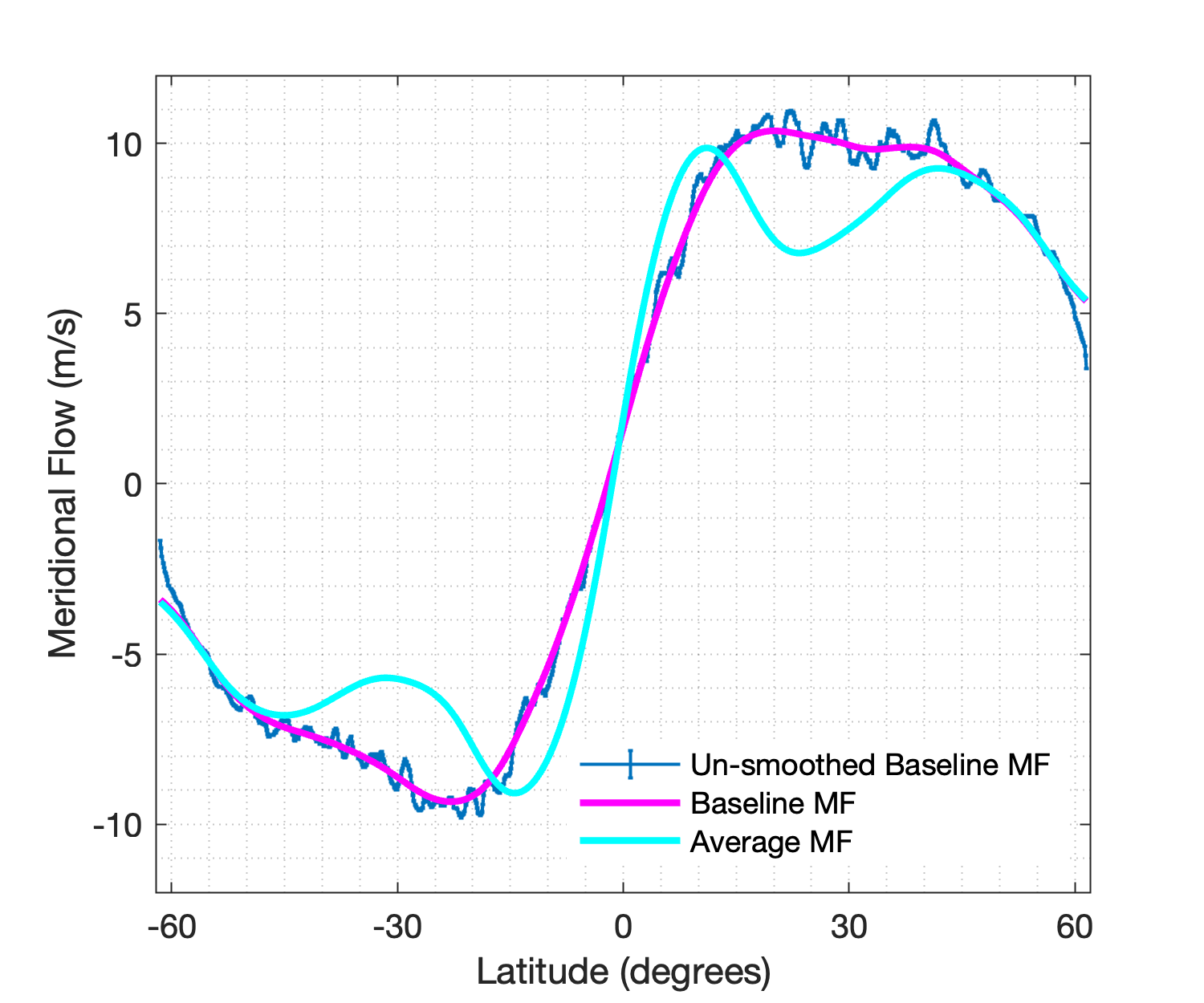}
    \caption{Baseline meridional flow is shown in blue. Statistical errors are included in the thickness of the blue curve. The smoothed baseline profile (magenta) and the average meridional flow from everywhere (cyan) are obtained by using a $15\degr$ FWHM moving Gaussian average.}
    \label{fig:baseline_mf}
\end{figure}

\subsection{Meridional Flow}

Here, we describe how the near-surface meridional flow can be broken down into three major components.

\subsubsection{Baseline Meridional Flow}
The meridional flow calculated within different annuli (top row of Fig.\ \ref{fig:mf_bfly_annuli}) shows that the perturbations caused by active region inflows are the strongest within the first $5\degr$ and largely disappear $30\degr$ away from active region centroids. More data points are available for averaging as the annuli increase in radius, which reduces the standard error of the mean. Conversely, flows outside of a chosen radius (bottom row of Fig.\ \ref{fig:mf_bfly_annuli}) show that the solar-cycle-scale variations gradually diminish as larger areas around active regions are left out from the calculation of meridional flow. For the cases where larger than $15\degr$ neighborhoods are excluded, we start running out of data points from the active latitudes (last three panels in the bottom row of Fig.~\ref{fig:mf_bfly_annuli}).

 Fig.\ \ref{fig:mf_curves} shows the temporal averages of meridional flow (in four bins of time, 31 Carrington Rotations each, during Sunspot Cycle 24) as a function of distance away from active region centroids. 
The curves in the top panel indicate that the meridional flow profile converges to the curve plotted in magenta as we look farther away from active regions. \textbf{We designate this magenta curve, shown in detail in Fig.\ \ref{fig:baseline_mf}, as the baseline meridional flow. It is calculated as the mean flow obtained from more than $30\degr$ 
(heliographic) and $48$ hours away from any active region. The profile is averaged over the entire period of Cycle 24, and is further smoothed with a $15\degr$ FWHM Gaussian running mean. } As indicated by the second row of Fig.\ \ref{fig:mf_bfly_annuli}, this baseline meridional flow profile is heavily reliant on meridional flow close to sunspot minima due to missing data during the sunspot maximum (owing to the high density of active regions, there is a significant chunk of latitudes that are never $30\degr$ away from all active regions during sunspot maximum). The similarly smoothed temporal average of meridional flow from all data differs from the baseline meridional flow profile due to contamination from inflows around active regions and is plotted in cyan for comparison. The bottom row of Fig.\ \ref{fig:mf_curves} shows the corresponding residual meridional flow profiles after the magenta curve is subtracted as background.

\subsubsection{AR contribution to Meridional Flow}

Figs.\ \ref{fig:mf_bfly_annuli} and \ref{fig:mf_curves} indicate a few key features: 1) the meridional flow farther away from active regions gradually approaches the baseline profile (shown in magenta); 2) the quadrupolar nature of the residual flow (after subtraction of the baseline meridional flow), along with decrease in perturbation amplitude as a function of distance away from active regions, are clear indicators of an inflow toward active region centroids as the source of these perturbations; 3) the area of influence of active regions stretches out up to $30\degr$ away from active region centroids; 4) the temporal average of meridional flow everywhere on the solar disk during Cycle 24 (cyan curve) is significantly influenced by inflows around active regions. Although this influence is diluted due to averaging over all annuli and background flow alike, it still makes a significant impact, at $\sim 4$~m~s$^{-1}$ level around $\pm 30\degr$ latitude. Thus, if one goes by the temporally averaged meridional flow taken from everywhere on the solar disk (cyan curve in Fig.\ \ref{fig:mf_curves}), one would overestimate the meridional flow amplitude at low latitudes and underestimate the amplitude at high latitudes due to contamination from inflows around active regions, and this error would be proportional to the strength of the sunspot activity during the averaging period. This could explain some of the differences between historically reported profiles of meridional flow from different solar cycles and even different phases within those solar cycles.

There is also an apparent change in the quadrupolar inflow signatures centered around activity belts as a function of the solar cycle. In the first three panels of the bottom row of Fig.\ \ref{fig:mf_curves}, the high-latitude wing of the inflow remains nearly the same magnitude and width (in latitude). On the other hand, the low-latitude wing of inflow gets squeezed toward the equator reducing its latitudinal width. This apparent temporal change could be due to a cross-equatorial overlap between neighboring active regions as the sunspot activity belt migrates to lower latitudes. This also means that if there is a hemispheric asymmetry in the strength of the sunspot number, it could lead to a cross-equatorial flow over the lifetime of the asymmetry. See, for example, the period between around 2014 -- 2015 in Fig.\ \ref{fig:mf_third} when the southern hemisphere was way more active than the northern hemisphere (the hemispheric asymmetry in sunspot numbers can be seen at: \url{https://www.sidc.be/silso/monthlyhemisphericplot}). During this time, the low-latitude inflow band in the northern hemisphere appears very weak, even broken as the active regions in the southern hemisphere pull inflows across the equator, thus diminishing the counterpart low latitude inflow band in the northern hemisphere.

\subsubsection{Residual Solar Cycle Scale Trend in Meridional Flow}
\begin{figure}
    \centering
    \includegraphics[width=\textwidth]{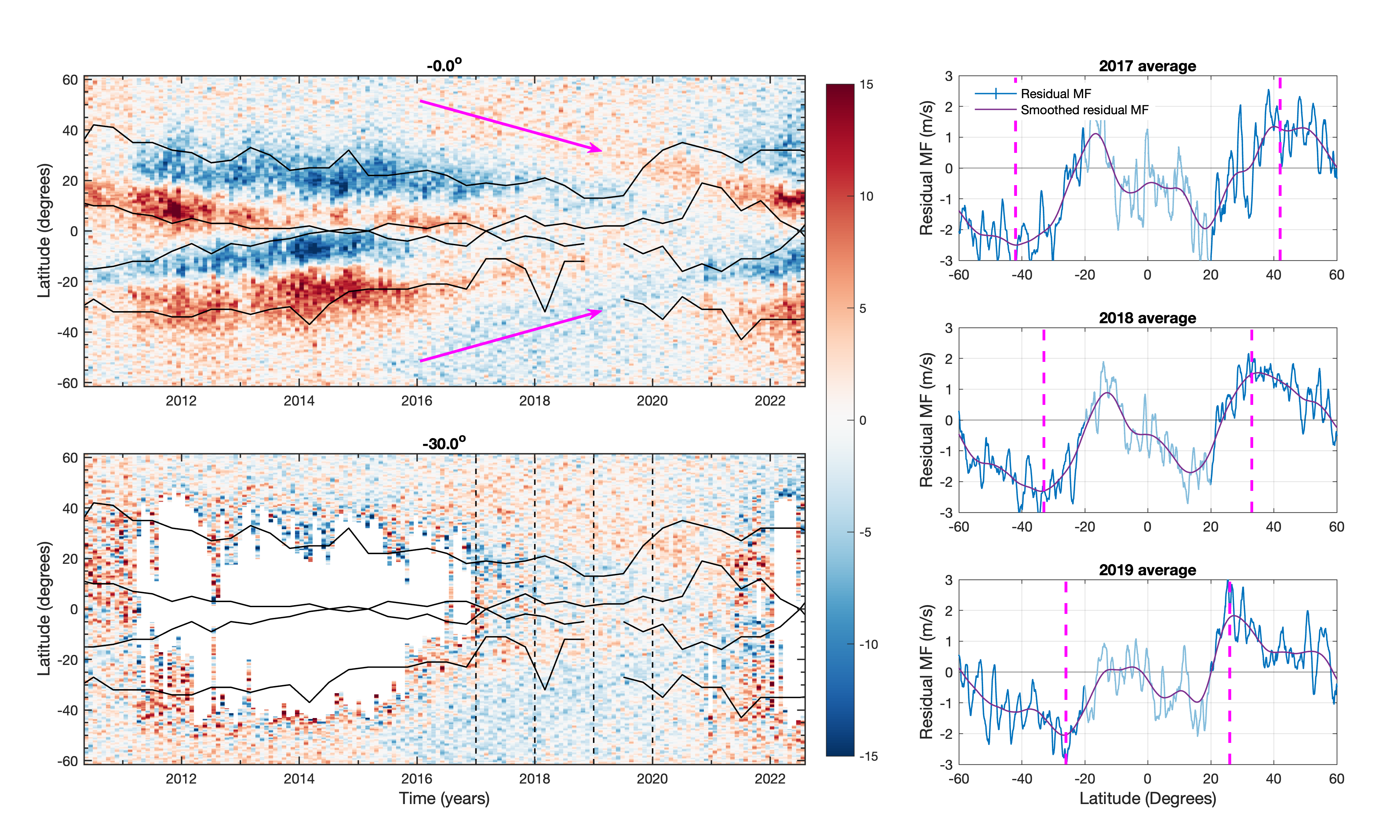}
    \caption{Top left: Residual meridional flow (after baseline flow subtraction) measured everywhere averaged over all Carrington rotations shows strong inflows into activity belts and weak bands of faster meridional flow migrating toward the equator near the end of Solar Cycle 24 (marked by magenta arrows). Bottom left: Residual meridional flow farther than $30\degr$ around active regions still shows the weak equatorward migrating bands marked by magenta arrows in the top panel. Right: Meridional flow farther than $30\degr$ from active regions averaged over the years 2017 (top), 2018 (middle) and 2019 (bottom) shows that the amplitude of these bands is $\sim$2~m~s$^{-1}$. The locations of the equatorward propagating bands in each year are marked by dashed magenta lines.}
    \label{fig:mf_third}
\end{figure}

Knowing the baseline meridional flow profile now enables us to create a butterfly diagram of the residual meridional flow which is dominated by active region inflows shown in the top row of Fig.\ \ref{fig:mf_third}. This plot, however, also shows weak bands of faster-than-average  meridional flow that originate in the year 2016 at high latitudes ($>50\degr$) and propagate toward the equator (marked by magenta arrows). This is a solar-cycle-scale variation that appears outside of the activity belts and appears invariable irrespective of the area around active regions excluded from the calculation of meridional flow (the bottom row of Fig.\ \ref{fig:mf_third} shows the $30\degr$ excluded residual meridional flow with the same bands). Averaging the residual meridional flow over the years 2017, 2018 and 2019 (time period between black dashed lines in the bottom panel, the averages are shown in the right panel of Fig.\ \ref{fig:mf_third}) shows that the amplitude of these bands is $\sim \pm 2$~m~s$^{-1}$ at $31\degr$ latitude in the southern hemisphere whereas it is $\sim \pm 1.5$~m~s$^{-1}$ at $31\degr$ latitude in the northern hemisphere in 2018. These bands of faster than baseline meridional flow propagate towards the equator in time with their peak amplitude appearing at latitudes $42\degr$, $33\degr$ and $26\degr$ in years 2017, 2018 and 2019 respectively. Thus, apart from the constant baseline meridional flow profile and short-lived inflows around active regions, meridional flow seems to have a third component that is weaker than the other two, varies over timescales of the solar cycle and it is clearly visible in quiet-Sun regions between two cycles.

This component of meridional flow has a bipolar signature in latitude rather than a quadrupolar signature like active region inflows. It may be indicative of variations in the true background meridional flow possibly arising due to the global dynamo or the appearance of ephemeral regions at high latitudes prior to the beginning of a cycle \citep{1985AuJPh..38..875H}. It also matches the behavior seen in several other observable properties on the solar surface like torsional oscillation, X-ray bright points, etc. that indicate an extended solar cycle \citep{1988Natur.333..748W,2014ApJ...792...12M}.

Given that this smaller flow perturbation is not spatio-temporally linked to active regions, we cannot truly separate it from background meridional flow. We, therefore, construct the baseline meridional flow as an average of the background meridional flow ($>30\degr$ and 48 hours away from active regions) over the entire cycle to dilute its effect on the baseline meridional flow.

Note here that this solar-cycle-scale trend in meridional flow can also be seen quite clearly prior to the subtraction of the baseline meridional flow in the bottom panels of Fig.\ \ref{fig:mf_bfly_annuli}. Thus, its existence and its equatorward propagation is a robust feature and independent from variations in the algorithm used to calculate the baseline meridional flow.

%%%%%%%%%%

\subsection{Torsional Oscillation}

Torsional oscillation is the residual of the rotation rate of the Sun left after subtracting a background profile. Usually this background profile is chosen as the temporal average over a solar cycle. As torsional oscillation is a flow in/against the direction of rotation, inflows around active regions should only affect the torsional oscillation if they are asymmetric around active regions. Thus, if the East-West inflows around active regions are symmetric, we should not see any variation in torsional oscillation pattern as a function of distance away from active regions.

\begin{figure*}
    \centering
   \includegraphics[width=\textwidth]{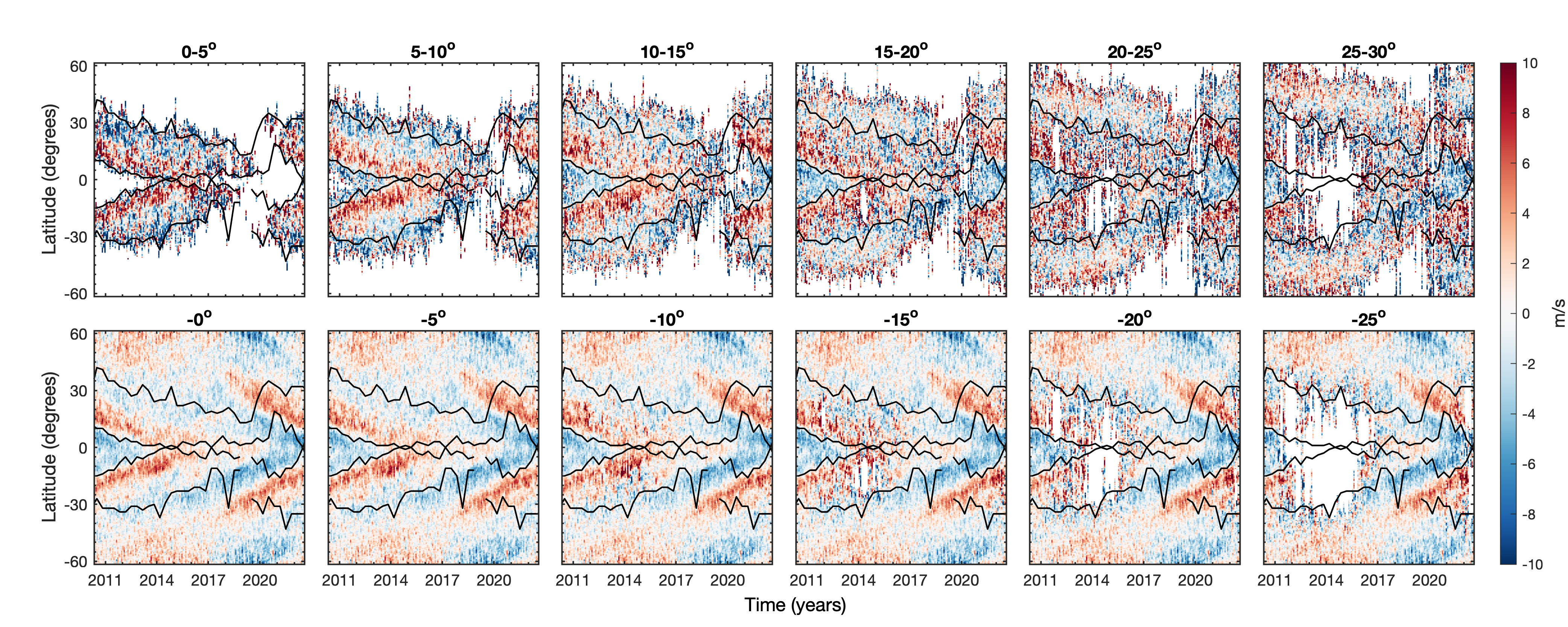}
    \caption{Similar to Fig.\ \ref{fig:mf_bfly_annuli}, but for torsional oscillation within several annuli around active regions in the top row and torsional oscillation outside excluded regions around sunspots in the bottom row. }
    \label{fig:dr_bfly_annuli}
\end{figure*}

Fig.\ \ref{fig:dr_bfly_annuli} shows the torsional oscillation pattern after the subtraction of temporal average of rotation rate at each latitude for the 10.83 years of data we have (nearly a full sunspot cycle) in the same manner as Fig.\ \ref{fig:mf_bfly_annuli}. It indicates that except within $5\degr$ from active region flux centroids, the torsional oscillation pattern does not significantly vary as a function of distance away from active regions i.e. columns 2 to 6 of Fig.\ \ref{fig:dr_bfly_annuli} are visually similar. The variation in rotation rate within the first $5\degr$ annulus is associated with sunspots rotating at a slightly different rate than the near-surface layers, a phenomenon well known in the solar physics community \citep{2003PhDT.........9G}. Overall, the East-West inflows into active regions cancel out quite well and the torsional oscillation pattern appears to be omnipresent in the near-surface layer of the Sun.  

\begin{figure*}
    \centering
    \includegraphics[width=\textwidth]{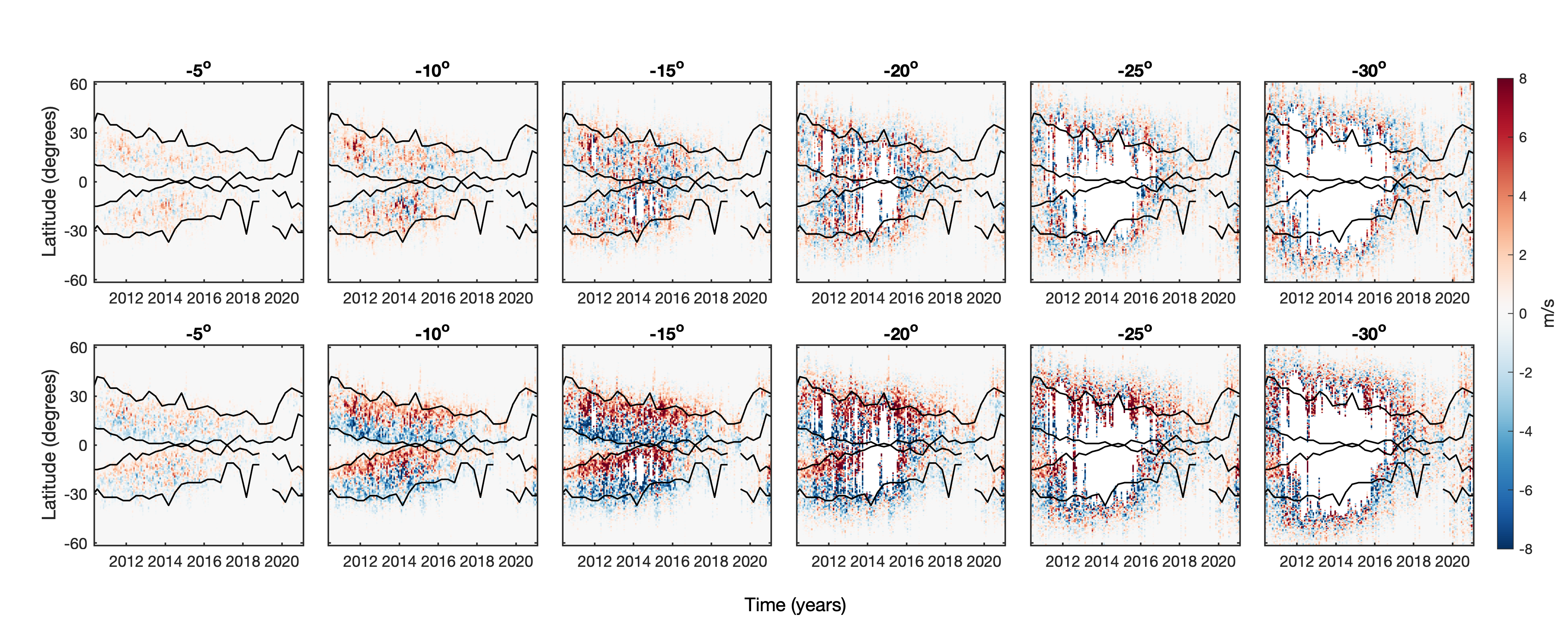}
    \caption{\textit{Top row:} The residual torsional oscillation pattern calculated after removing the torsional oscillation pattern everywhere (the one shown in the first panel of the bottom row in Fig.\ \ref{fig:dr_bfly_annuli}) from the measurements at other cutout sizes (size of area around ARs excluded). The title of each panel represents the size of cutout excluded around ARs. There is no clear difference in torsional oscillation pattern as a function of cutout size as the residuals appear quite random. \textit{Bottom row:} The residual meridional flow after subtracting meridional flow measured everywhere (bottom left panel of Fig.\ \ref{fig:mf_bfly_annuli}) from meridional flow measurements at other cutout sizes. This indicates a clear solar cycle scale pattern in meridional flow as a function of cutout size.}
    \label{fig:error_analysis}
\end{figure*}

In order to test that the variation in torsional oscillation pattern with respect to the size of excluded area around active regions is truly insignificant, we plot the residual of torsional oscillation shown in the bottom row of Fig.\ \ref{fig:dr_bfly_annuli} after removing the torsional oscillation measured without any exclusion of ARs (i.e. first panel in the bottom row of Fig.\ \ref{fig:dr_bfly_annuli}) in Fig.\ \ref{fig:error_analysis}. Similarly, residual meridional flow maps were created from the bottom row of Fig.\ \ref{fig:mf_bfly_annuli} for comparison by removing meridional flow measured without the exclusion of any ARs. While the variation in meridional flow with cutout size (size of area around ARs excluded) is easily apparent in both Fig.\ \ref{fig:mf_bfly_annuli} and in Fig.\ \ref{fig:error_analysis}, the variation in torsional oscillation is neither apparent in Fig.\ \ref{fig:dr_bfly_annuli} nor in the top row of Fig.\ \ref{fig:error_analysis}. While there may be some weak large scale pattern to the residual torsional oscillation with cutout size less than $10\degr$, there is no evidence for a large scale trend for cutout sizes greater than $10\degr$.

%\begin{figure*}
%    \centering
%    \includegraphics[width=\textwidth]{to_bfly_diff.png}
%    \caption{The difference in torsional oscillation %pattern compared to torsional oscillation everywhere.}
%    \label{fig:to_bfly_diff}
%\end{figure*}

The choice of a background profile for torsional oscillation, therefore, is trickier than for meridional flow. The fact that the torsional oscillation pattern does not go away when active regions are excluded implies that it is also a part of the background East-West flow. On top of this, the torsional oscillation pattern varies with time making it impossible to construct a simple background East-West flow profile that is independent of time. However, active regions do not seem to affect torsional oscillation farther than $10\degr$ away (see Fig.\ \ref{fig:error_analysis}). Thus, the torsional oscillation pattern measured by excluding $10\degr$ (heliographic) around active region centroids (on top of the mean differential rotation profile subtracted to reveal torsional oscillation in the first place) seems to be a reasonably good estimate of the background rotation rate of the Sun and can be used to
isolate the local East-West flows in the vicinity of active regions. This background torsional oscillation profile is shown in the third panel of the bottom row of Fig.\ \ref{fig:dr_bfly_annuli}). It may be re-iterated that choosing a constant temporal mean of the differential rotation profile that only varies with latitude would not represent the background near-surface rotation rate well.

%pattern $10\degr$ away along with its latitudinal and temporal variation as shown in the third panel of the bottom row of Fig.\ \ref{fig:dr_bfly_annuli} should be chosen as the background. Choosing a constant mean differential rotation profile that only varies with latitude  would not represent the background near-surface rotation rate well.

% \begin{figure}
%     \centering
%     \includegraphics[width = 18 cm]{../figures3/ARcut_MF_temporal_dependence_isolation20.png}
%     \caption{Same as above figure, except that residual MF profiles after subtracting the black curve(background MF beyond $20\degr$ away from AR centroids) are shown. Within $5\degr$ of AR centroids, the inflows seem to peak $\sim 12m/s$, while they reduce to $\sim 8 m/s$ in the $5\degr-10\degr$ annuli. The inflows further decrease to around $\sim 4m/s$ in the $10\degr-15\degr$ annuli while they are barely noticeable in the $15\degr-20\degr$ annuli.}
%     \label{fig:mf_cuts_temporal}
% \end{figure}
% 
% \begin{figure}
%     \centering
%     \includegraphics[width = 18 cm]{../figures3/ARcut_MF_latitude_dependence_isolation20.png}
%     \caption{Same as above figure, except that residual MF profiles after subtracting the black curve(background MF beyond $20\degr$ away from AR centroids) are shown. Within $5\degr$ of AR centroids, the inflows seem to peak $\sim 12m/s$, while they reduce to $\sim 8 m/s$ in the $5\degr-10\degr$ annuli. The inflows further decrease to around $\sim 4m/s$ in the $10\degr-15\degr$ annuli while they are barely noticeable in the $15\degr-20\degr$ annuli.}
%     \label{fig:mf_cuts_lat}
% \end{figure}

\section{Discussion} \label{sec:discussion}

We need to point out that the inflows, as measured near the active regions, may suffer from systematics in the local helioseismology techniques. It is well known that the interaction between magnetic field and helioseismic $p$-mode waves is rather complicated in sunspot regions, and disentangling these interactions for an accurate inversion of subsurface flows in active regions is also complex. The travel-time differences measured between $p$-mode waves traveling into and out from sunspots are understandably affected by the presence of magnetic fields as well as the Wilson Depression. It is often believed that these effects add apparent inward flows to the inverted subsurface flows near active regions, although it is unclear what fraction of the inverted inflows corresponds to true plasma flow and what fraction is due to the systematic effect in the helioseismic analysis. Nevertheless, our current approach of removing flows up to $30\degr$ away from active region centroids %removes both of the inflow fractions
should remove the true inflow and systematics alike, and thus circumvents the ambiguity in the composition of these inflows.

%Historical meridional flow variation found during the solar cycle is exaggerated because it includes active region inflows. After taking those out, the background meridional flow doesn't change much throughout a solar cycle.
Historical measurements of meridional flow variation found during solar cycles \citep[e.g.][]{1993SoPh..147..207K,1998ApJ...508L.105B,2003PhDT.........9G,2004ApJ...603..776Z,2010Sci...327.1350H,2011ApJ...729...80H, 2014ApJ...789L...7Z,2021ApJ...917..100M} generally did not exclude the contributions from active region inflows. These active region inflows can result in the peak meridional flow being measured at different latitudes during different phases of the solar cycle (see Fig.\ \ref{fig:mf_curves}). This knowledge can be used to put historical observations of meridional flow from different time periods into perspective by enabling meaningful comparison. %However, it is useful to stress here that the taken-out ``inflows" may not all be the plasma inflows, but are likely a combination of plasma flows and systematics in the local helioseismic techniques.

Several papers in the past have studied flows within active regions versus flows in quiet-Sun regions. We split them up into these groups: 1) \citet{2008ApJ...680L.161S} 2) \citet{2008SoPh..252..235G} 3) \citet{2016A&A...590A.130L,2016ApJ...819..106B,2019ApJ...873...94B,2022A&A...664A.189P} and 4) \citet{2020SoPh..295...47K}. They have all used different criteria to differentiate between active regions and quiet-Sun regions on the Sun. One common aspect of active region selection in all these studies is that the selection is based on an average magnetic field strength from a synoptic map. Studies in groups 1,2 and 3 place a rectangular box around the centers of peak flux in a synoptic map and carve out an area within 7.5 to 12 degrees around the peak as the active region neighborhood. All areas that do not fall into any active region neighborhood are automatically classified as quiet-Sun regions. Such a selection methodology does well in identifying the locations of active regions, but is not well suited to capture the full extent of inflows around active regions as it assumes the inflows do not stretch beyond 12 degrees away from active regions. Actually, inflows are expected to be nearly zero at the flux centroids of active regions and are expected to have a peak outside of active regions. According to our results, these studies would have ended up with a significant variation in meridional flow arising from inflows that stretch beyond 12 degrees in the quiet-Sun regions like in the panel marked $-10\degr$ in Fig.\ \ref{fig:mf_bfly_annuli} which shows meridional measured farther than $10\degr$ away from active region flux centroids. Thus, the background “quiet-Sun flows” that were subtracted to analyze flows around active regions in these studies could have been contaminated significantly with inflows around ARs. Besides, a synoptic map is an average of the region passing the central meridian, which does not reflect the temporal evolution of the AR. This may be a factor that could influence their results, too.

\citet{2020SoPh..295...47K} used a slightly different method for differentiation between active and quiet-Sun regions. They relied on the MAI (magnetic activity index) which is a quantity that is calculated as an average over $\pm 7.5\degr$ regions (i.e. $15\degr$ size tiles). They choose different thresholds $\mathrm{MAI} <3.7$ for quiet-Sun regions and $\mathrm{MAI} > 8.4$ G for identifying active regions. This makes their separation between active and quiet-Sun regions better, but restriction of the spatial extent to $\pm 7.5\degr$ would still result in the full extent of inflows around active regions not being captured in their active region subset. This could explain why all these studies find only slightly attenuated variation in meridional flow in their quiet-Sun regions. We, on the other hand, by not assuming a set value of area of influence around active regions are able to first establish that inflows around active regions can stretch up to $30\degr$ away and use this knowledge to separate active region neighborhoods (within $30\degr$ and 48 hours away) from quiet-Sun regions.

Another difference between the active region selection criteria in previous studies and our current study is that some of the previous studies limited their active region selection to capture active regions with net magnetic flux $> 10^{21}$ Mx \citep{2019ApJ...873...94B,2022A&A...664A.189P}. How this relates to the USFLUX we use is unclear. We, on the other hand do not make any such assumption and use all detected active regions that are a part of the SHARP data series. We find that $39.55\%$ of SHARP instances are below $10^{21}$ Mx in USFLUX and are included in our study while they could probably be excluded from previous studies.

Moreover, to be super conservative, we also exclude 30 degree radius neighborhoods around the expected locations of all active regions (SHARPs) 48 hours prior to their first record and 48 hours post their last record from quiet-Sun regions. All the above differences imply that previous studies had a less strict criterion for the separation of active and quiet-Sun regions that lead to some part of the inflows around active regions showing up in both their active and quiet-Sun subsets. By making the selection criterion much more conservative, we are able to remove all signs of inflows around active regions from the quiet-Sun regions and reveal the actual variation in quiet-Sun regions that appears between Cycles 24 and 25 at high latitudes in Fig.\ \ref{fig:mf_third}. This mysterious bipolar (in latitude) signal (marked by magenta arrows in Fig.\ \ref{fig:mf_third}) appears to have a fundamentally different origin disparate from active regions and could be indicative of an extended solar cycle, or a meridional flow variation driven directly by the solar dynamo. This could be an exciting new phenomenon, but needs to be confirmed by other velocity measurement techniques. The presence of this solar-cycle-scale variation in meridional flow which, to our knowledge, has not been reported before (after a careful omission of active regions) can be used as a new constraint on solar dynamo models.

Our results explain why \citet{2010ApJ...720.1030C} were able to reproduce most variations in of meridional flow by simply adding inflows towards activity belts on top of a constant background meridional flow profile. This is due to the fact that the quiet-Sun meridional flow only shows a very weak solar-cycle-scale variation ($\sim 2 m/s$) after active region inflows and the baseline meridional flow are removed as shown in Fig.\ \ref{fig:mf_third}. \citet{2012A&A...548A..57C} later showed that such inflows around active regions can modulate the strength of the polar field built up at the end of a sunspot cycle which is a strong proxy for the strength of the upcoming sunspot cycle \citep{2013ApJ...767L..25M}. The effect manifests as a correlation between the net flux in the poleward magnetic ``surges'' and the inflow velocity \citep{2014ApJ...789L...7Z,2015ApJ...798..114S}.

 The fact that active region inflows stretch so far away from active region flux centroids also implies that these inflows could drive cross-equatorial flows when there is an imbalance of active regions between the northern and southern hemispheres. The existence of cross-equatorial flows throughout the solar cycle can enhance cancellation of magnetic flux across the equator and thus have a direct impact on the build up of polar fields and consequently the amplitude of the next cycle.

In our analysis, active region inflows in the direction of rotation largely cancel out and do not significantly contribute to the torsional oscillation pattern. This indicates that torsional oscillation could be a global phenomenon and might be unrelated to flow perturbations driven by active regions at the solar surface. This is an important constraint for solar dynamo and surface flux transport models. The fact that torsional oscillation is unrelated to inflows around active regions may be evidence against the notion of \citet{2003SoPh..213....1S} who proposed that inflows into activity belts would cause torsional oscillation by driving geostrophic flows.

The methodology used in this study to separate active regions from quiet-Sun regions 
is also applicable to velocity maps obtained from other measurement techniques to infer background flow profiles and improve measurements of flows around other features on the Sun. In the near future, the constant baseline meridional flow profile and the temporally variable baseline torsional oscillation profile identified as the near-surface background flow profile in this study will be used to correct velocity maps in the vicinity of active regions in order to analyze the flows around active regions in greater detail in the proper frame of reference.

All data required to re-create Figs.\ 2 through 7 are available at the Standford Digital Repository at \url{https://purl.stanford.edu/ph433nq0725} and \dataset[DOI: 10.25740/ph433nq0725]{https://doi.org/10.25740/ph433nq0725} which includes the baseline meridional flow, meridional flow as a function of distance away from active regions and torsional oscillation as a function of distance away from active regions.

\acknowledgements{This work is supported by NASA Grant 80NSSC20K0184 to Stanford University, with subawards to the University of Hawaii and to Lockheed Martin. Data used in this work are courtesy of NASA/SDO and the HMI science team.}

%%%%%%%%%%%%%%%%%%%%%%%%---FIGURES---%%%%%%%%%%%%%%%%%%%%%%%%%%%%%%%%%%

%\appendix

\acknowledgements

\bibliographystyle{apj}
\bibliography{references}

\begin{thebibliography}{}
\expandafter\ifx\csname natexlab\endcsname\relax\def\natexlab#1{#1}\fi
\providecommand{\url}[1]{\href{#1}{#1}}
\providecommand{\dodoi}[1]{doi:~\href{http://doi.org/#1}{\nolinkurl{#1}}}
\providecommand{\doeprint}[1]{\href{http://ascl.net/#1}{\nolinkurl{http://ascl.net/#1}}}
\providecommand{\doarXiv}[1]{\href{https://arxiv.org/abs/#1}{\nolinkurl{https://arxiv.org/abs/#1}}}

\bibitem[{{Bobra} {et~al.}(2014){Bobra}, {Sun}, {Hoeksema}, {Turmon}, {Liu},
  {Hayashi}, {Barnes}, \& {Leka}}]{2014SoPh..289.3549B}
{Bobra}, M.~G., {Sun}, X., {Hoeksema}, J.~T., {et~al.} 2014, \solphys, 289,
  3549, \dodoi{10.1007/s11207-014-0529-3}

\bibitem[{{Braun}(2016)}]{2016ApJ...819..106B}
{Braun}, D.~C. 2016, \apj, 819, 106, \dodoi{10.3847/0004-637X/819/2/106}

\bibitem[{{Braun}(2019)}]{2019ApJ...873...94B}
---. 2019, \apj, 873, 94, \dodoi{10.3847/1538-4357/ab04a3}

\bibitem[{{Braun} \& {Fan}(1998)}]{1998ApJ...508L.105B}
{Braun}, D.~C., \& {Fan}, Y. 1998, \apjl, 508, L105, \dodoi{10.1086/311727}

\bibitem[{{Cameron} \& {Sch{\"u}ssler}(2010)}]{2010ApJ...720.1030C}
{Cameron}, R.~H., \& {Sch{\"u}ssler}, M. 2010, \apj, 720, 1030,
  \dodoi{10.1088/0004-637X/720/2/1030}

\bibitem[{{Cameron} \& {Sch{\"u}ssler}(2012)}]{2012A&A...548A..57C}
---. 2012, \aap, 548, A57, \dodoi{10.1051/0004-6361/201219914}

\bibitem[{{Carrington}(1858)}]{1858MNRAS..19....1C}
{Carrington}, R.~C. 1858, \mnras, 19, 1, \dodoi{10.1093/mnras/19.1.1}

\bibitem[{{Chen} \& {Zhao}(2017)}]{2017ApJ...849..144C}
{Chen}, R., \& {Zhao}, J. 2017, \apj, 849, 144,
  \dodoi{10.3847/1538-4357/aa8eec}

\bibitem[{{Couvidat} {et~al.}(2016){Couvidat}, {Schou}, {Hoeksema}, {Bogart},
  {Bush}, {Duvall}, {Liu}, {Norton}, \& {Scherrer}}]{2016SoPh..291.1887C}
{Couvidat}, S., {Schou}, J., {Hoeksema}, J.~T., {et~al.} 2016, \solphys, 291,
  1887, \dodoi{10.1007/s11207-016-0957-3}

\bibitem[{{Duvall}(1979)}]{1979SoPh...63....3D}
{Duvall}, T.~L., J. 1979, \solphys, 63, 3, \dodoi{10.1007/BF00155690}

\bibitem[{{Duvall} {et~al.}(1993){Duvall}, {Jefferies}, {Harvey}, \&
  {Pomerantz}}]{1993Natur.362..430D}
{Duvall}, T.~L., J., {Jefferies}, S.~M., {Harvey}, J.~W., \& {Pomerantz}, M.~A.
  1993, \nat, 362, 430, \dodoi{10.1038/362430a0}

\bibitem[{{Gizon}(2003)}]{2003PhDT.........9G}
{Gizon}, L. 2003, PhD thesis, STANFORD UNIVERSITY

\bibitem[{{Gizon}(2004)}]{2004SoPh..224..217G}
---. 2004, \solphys, 224, 217, \dodoi{10.1007/s11207-005-4983-9}

\bibitem[{{Gonz{\'a}lez Hern{\'a}ndez} {et~al.}(2008){Gonz{\'a}lez
  Hern{\'a}ndez}, {Kholikov}, {Hill}, {Howe}, \& {Komm}}]{2008SoPh..252..235G}
{Gonz{\'a}lez Hern{\'a}ndez}, I., {Kholikov}, S., {Hill}, F., {Howe}, R., \&
  {Komm}, R. 2008, \solphys, 252, 235, \dodoi{10.1007/s11207-008-9264-y}

\bibitem[{{Harvey} {et~al.}(1996){Harvey}, {Hill}, {Hubbard}, {Kennedy},
  {Leibacher}, {Pintar}, {Gilman}, {Noyes}, {Title}, {Toomre}, {Ulrich},
  {Bhatnagar}, {Kennewell}, {Marquette}, {Patron}, {Saa}, \&
  {Yasukawa}}]{1996Sci...272.1284H}
{Harvey}, J.~W., {Hill}, F., {Hubbard}, R.~P., {et~al.} 1996, Science, 272,
  1284, \dodoi{10.1126/science.272.5266.1284}

\bibitem[{{Harvey}(1985)}]{1985AuJPh..38..875H}
{Harvey}, K.~L. 1985, Australian Journal of Physics, 38, 875,
  \dodoi{10.1071/PH850875}

\bibitem[{{Hathaway} \& {Rightmire}(2010)}]{2010Sci...327.1350H}
{Hathaway}, D.~H., \& {Rightmire}, L. 2010, Science, 327, 1350,
  \dodoi{10.1126/science.1181990}

\bibitem[{{Hathaway} \& {Rightmire}(2011)}]{2011ApJ...729...80H}
---. 2011, \apj, 729, 80, \dodoi{10.1088/0004-637X/729/2/80}

\bibitem[{{Hill}(1988)}]{1988ApJ...333..996H}
{Hill}, F. 1988, \apj, 333, 996, \dodoi{10.1086/166807}

\bibitem[{{Howard} \& {LaBonte}(1980)}]{1980ApJ...239L..33H}
{Howard}, R., \& {LaBonte}, B.~J. 1980, \apjl, 239, L33, \dodoi{10.1086/183286}

\bibitem[{{Howe} {et~al.}(2013){Howe}, {Christensen-Dalsgaard}, {Hill}, {Komm},
  {Larson}, {Rempel}, {Schou}, \& {Thompson}}]{2013ApJ...767L..20H}
{Howe}, R., {Christensen-Dalsgaard}, J., {Hill}, F., {et~al.} 2013, \apjl, 767,
  L20, \dodoi{10.1088/2041-8205/767/1/L20}

\bibitem[{{Howe} {et~al.}(2009){Howe}, {Christensen-Dalsgaard}, {Hill}, {Komm},
  {Schou}, \& {Thompson}}]{2009ApJ...701L..87H}
---. 2009, \apjl, 701, L87, \dodoi{10.1088/0004-637X/701/2/L87}

\bibitem[{{Howe} {et~al.}(2018){Howe}, {Hill}, {Komm}, {Chaplin}, {Elsworth},
  {Davies}, {Schou}, \& {Thompson}}]{2018ApJ...862L...5H}
{Howe}, R., {Hill}, F., {Komm}, R., {et~al.} 2018, \apjl, 862, L5,
  \dodoi{10.3847/2041-8213/aad1ed}

\bibitem[{{Howe} {et~al.}(2000){Howe}, {Komm}, \& {Hill}}]{2000SoPh..192..427H}
{Howe}, R., {Komm}, R., \& {Hill}, F. 2000, Solar Physics, 192, 427,
  \dodoi{10.1023/A:1005202805104}

\bibitem[{{Howe} {et~al.}(2004){Howe}, {Rempel}, {Christensen-Dalsgaard},
  {Hill}, {Komm}, {Schou}, \& {Thompson}}]{2004ESASP.559..468H}
{Howe}, R., {Rempel}, M., {Christensen-Dalsgaard}, J., {et~al.} 2004, in ESA
  Special Publication, Vol. 559, SOHO 14 Helio- and Asteroseismology: Towards a
  Golden Future, ed. D.~{Danesy}, 468

\bibitem[{{Komm} {et~al.}(2020){Komm}, {Howe}, \& {Hill}}]{2020SoPh..295...47K}
{Komm}, R., {Howe}, R., \& {Hill}, F. 2020, \solphys, 295, 47,
  \dodoi{10.1007/s11207-020-01611-5}

\bibitem[{{Komm}(1994)}]{1994SoPh..149..417K}
{Komm}, R.~W. 1994, \solphys, 149, 417, \dodoi{10.1007/BF00690628}

\bibitem[{{Komm} {et~al.}(1993){Komm}, {Howard}, \&
  {Harvey}}]{1993SoPh..147..207K}
{Komm}, R.~W., {Howard}, R.~F., \& {Harvey}, J.~W. 1993, \solphys, 147, 207,
  \dodoi{10.1007/BF00690713}

\bibitem[{{LaBonte} \& {Howard}(1982{\natexlab{a}})}]{1982SoPh...75..161L}
{LaBonte}, B.~J., \& {Howard}, R. 1982{\natexlab{a}}, Solar Physics, 75, 161,
  \dodoi{10.1007/BF00153469}

\bibitem[{{LaBonte} \& {Howard}(1982{\natexlab{b}})}]{1982SoPh...80..361L}
---. 1982{\natexlab{b}}, \solphys, 80, 361, \dodoi{10.1007/BF00147982}

\bibitem[{{L{\"o}ptien} {et~al.}(2016){L{\"o}ptien}, {Birch}, {Duvall},
  {Gizon}, \& {Schou}}]{2016A&A...590A.130L}
{L{\"o}ptien}, B., {Birch}, A.~C., {Duvall}, T.~L., {Gizon}, L., \& {Schou}, J.
  2016, \aap, 590, A130, \dodoi{10.1051/0004-6361/201628112}

\bibitem[{{Mahajan} {et~al.}(2021){Mahajan}, {Hathaway}, {Mu{\~n}oz-Jaramillo},
  \& {Martens}}]{2021ApJ...917..100M}
{Mahajan}, S.~S., {Hathaway}, D.~H., {Mu{\~n}oz-Jaramillo}, A., \& {Martens},
  P.~C. 2021, \apj, 917, 100, \dodoi{10.3847/1538-4357/ac0a80}

\bibitem[{{McIntosh} \& {Leamon}(2017)}]{2017FrASS...4....4M}
{McIntosh}, S.~W., \& {Leamon}, R.~J. 2017, Frontiers in Astronomy and Space
  Sciences, 4, 4, \dodoi{10.3389/fspas.2017.00004}

\bibitem[{{McIntosh} {et~al.}(2014){McIntosh}, {Wang}, {Leamon}, {Davey},
  {Howe}, {Krista}, {Malanushenko}, {Markel}, {Cirtain}, {Gurman}, {Pesnell},
  \& {Thompson}}]{2014ApJ...792...12M}
{McIntosh}, S.~W., {Wang}, X., {Leamon}, R.~J., {et~al.} 2014, \apj, 792, 12,
  \dodoi{10.1088/0004-637X/792/1/12}

\bibitem[{{Mu{\~n}oz-Jaramillo} {et~al.}(2013){Mu{\~n}oz-Jaramillo},
  {Dasi-Espuig}, {Balmaceda}, \& {DeLuca}}]{2013ApJ...767L..25M}
{Mu{\~n}oz-Jaramillo}, A., {Dasi-Espuig}, M., {Balmaceda}, L.~A., \& {DeLuca},
  E.~E. 2013, \apj, 767, L25, \dodoi{10.1088/2041-8205/767/2/L25}

\bibitem[{NOAA(1874)}]{RGO}
NOAA. 1874, {Royal Greenwich Observatory – USAF/NOAA Sunspot Data}.
\newblock \url{www.solarcyclescience.com}

\bibitem[{{Poulier} {et~al.}(2022){Poulier}, {Liang}, {Fournier}, \&
  {Gizon}}]{2022A&A...664A.189P}
{Poulier}, P.~L., {Liang}, Z.~C., {Fournier}, D., \& {Gizon}, L. 2022, \aap,
  664, A189, \dodoi{10.1051/0004-6361/202243476}

\bibitem[{{Scherrer} {et~al.}(1995){Scherrer}, {Bogart}, {Bush}, {Hoeksema},
  {Kosovichev}, {Schou}, {Rosenberg}, {Springer}, {Tarbell}, {Title},
  {Wolfson}, {Zayer}, \& {MDI Engineering Team}}]{1995SoPh..162..129S}
{Scherrer}, P.~H., {Bogart}, R.~S., {Bush}, R.~I., {et~al.} 1995, \solphys,
  162, 129, \dodoi{10.1007/BF00733429}

\bibitem[{{Scherrer} {et~al.}(2012){Scherrer}, {Schou}, {Bush}, {Kosovichev},
  {Bogart}, {Hoeksema}, {Liu}, {Duvall}, {Zhao}, {Title}, {Schrijver},
  {Tarbell}, \& {Tomczyk}}]{2012SoPh..275..207S}
{Scherrer}, P.~H., {Schou}, J., {Bush}, R.~I., {et~al.} 2012, \solphys, 275,
  207, \dodoi{10.1007/s11207-011-9834-2}

\bibitem[{{Schou} {et~al.}(1998){Schou}, {Antia}, {Basu}, {Bogart}, {Bush},
  {Chitre}, {Christensen-Dalsgaard}, {di Mauro}, {Dziembowski}, {Eff-Darwich},
  {Gough}, {Haber}, {Hoeksema}, {Howe}, {Korzennik}, {Kosovichev}, {Larsen},
  {Pijpers}, {Scherrer}, {Sekii}, {Tarbell}, {Title}, {Thompson}, \&
  {Toomre}}]{1998ApJ...505..390S}
{Schou}, J., {Antia}, H.~M., {Basu}, S., {et~al.} 1998, \apj, 505, 390,
  \dodoi{10.1086/306146}

\bibitem[{{Snodgrass} {et~al.}(1984){Snodgrass}, {Howard}, \&
  {Webster}}]{1984SoPh...90..199S}
{Snodgrass}, H.~B., {Howard}, R., \& {Webster}, L. 1984, \solphys, 90, 199,
  \dodoi{10.1007/BF00153796}

\bibitem[{{Spruit}(2003)}]{2003SoPh..213....1S}
{Spruit}, H.~C. 2003, \solphys, 213, 1, \dodoi{10.1023/A:1023202605379}

\bibitem[{{Sun} {et~al.}(2015){Sun}, {Hoeksema}, {Liu}, \&
  {Zhao}}]{2015ApJ...798..114S}
{Sun}, X., {Hoeksema}, J.~T., {Liu}, Y., \& {Zhao}, J. 2015, \apj, 798, 114,
  \dodoi{10.1088/0004-637X/798/2/114}

\bibitem[{{Toomre} {et~al.}(2000){Toomre}, {Christensen-Dalsgaard}, {Howe},
  {Larsen}, {Schou}, \& {Thompson}}]{2000SoPh..192..437T}
{Toomre}, J., {Christensen-Dalsgaard}, J., {Howe}, R., {et~al.} 2000, \solphys,
  192, 437, \dodoi{10.1023/A:1005223308315}

\bibitem[{{Ulrich} {et~al.}(1988){Ulrich}, {Boyden}, {Webster}, {Snodgrass},
  {Padilla}, {Gilman}, \& {Shieber}}]{1988SoPh..117..291U}
{Ulrich}, R.~K., {Boyden}, J.~E., {Webster}, L., {et~al.} 1988, \solphys, 117,
  291, \dodoi{10.1007/BF00147250}

\bibitem[{{Vorontsov} {et~al.}(2002){Vorontsov}, {Christensen-Dalsgaard},
  {Schou}, {Strakhov}, \& {Thompson}}]{2002Sci...296..101V}
{Vorontsov}, S.~V., {Christensen-Dalsgaard}, J., {Schou}, J., {Strakhov},
  V.~N., \& {Thompson}, M.~J. 2002, Science, 296, 101,
  \dodoi{10.1126/science.1069190}

\bibitem[{{{\v{S}}vanda} {et~al.}(2008){{\v{S}}vanda}, {Kosovichev}, \&
  {Zhao}}]{2008ApJ...680L.161S}
{{\v{S}}vanda}, M., {Kosovichev}, A.~G., \& {Zhao}, J. 2008, \apjl, 680, L161,
  \dodoi{10.1086/589997}

\bibitem[{{Wilson} {et~al.}(1988){Wilson}, {Altrocki}, {Harvey}, {Martin}, \&
  {Snodgrass}}]{1988Natur.333..748W}
{Wilson}, P.~R., {Altrocki}, R.~C., {Harvey}, K.~L., {Martin}, S.~F., \&
  {Snodgrass}, H.~B. 1988, \nat, 333, 748, \dodoi{10.1038/333748a0}

\bibitem[{{Zhao} \& {Kosovichev}(2004)}]{2004ApJ...603..776Z}
{Zhao}, J., \& {Kosovichev}, A.~G. 2004, \apj, 603, 776, \dodoi{10.1086/381489}

\bibitem[{{Zhao} {et~al.}(2014){Zhao}, {Kosovichev}, \&
  {Bogart}}]{2014ApJ...789L...7Z}
{Zhao}, J., {Kosovichev}, A.~G., \& {Bogart}, R.~S. 2014, \apjl, 789, L7,
  \dodoi{10.1088/2041-8205/789/1/L7}

\bibitem[{{Zhao} {et~al.}(2012{\natexlab{a}}){Zhao}, {Nagashima}, {Bogart},
  {Kosovichev}, \& {Duvall}}]{2012ApJ...749L...5Z}
{Zhao}, J., {Nagashima}, K., {Bogart}, R.~S., {Kosovichev}, A.~G., \& {Duvall},
  Jr., T.~L. 2012{\natexlab{a}}, \apjl, 749, L5,
  \dodoi{10.1088/2041-8205/749/1/L5}

\bibitem[{{Zhao} {et~al.}(2012{\natexlab{b}}){Zhao}, {Couvidat}, {Bogart},
  {Parchevsky}, {Birch}, {Duvall}, {Beck}, {Kosovichev}, \&
  {Scherrer}}]{2012SoPh..275..375Z}
{Zhao}, J., {Couvidat}, S., {Bogart}, R.~S., {et~al.} 2012{\natexlab{b}},
  \solphys, 275, 375, \dodoi{10.1007/s11207-011-9757-y}

\end{thebibliography}

\appendix

\section{Systematic Corrections}\label{sec:systematics}

\subsection{Center-to-limb Correction}

\begin{figure}[ht]
    \centering
    \includegraphics[width = 6 cm]{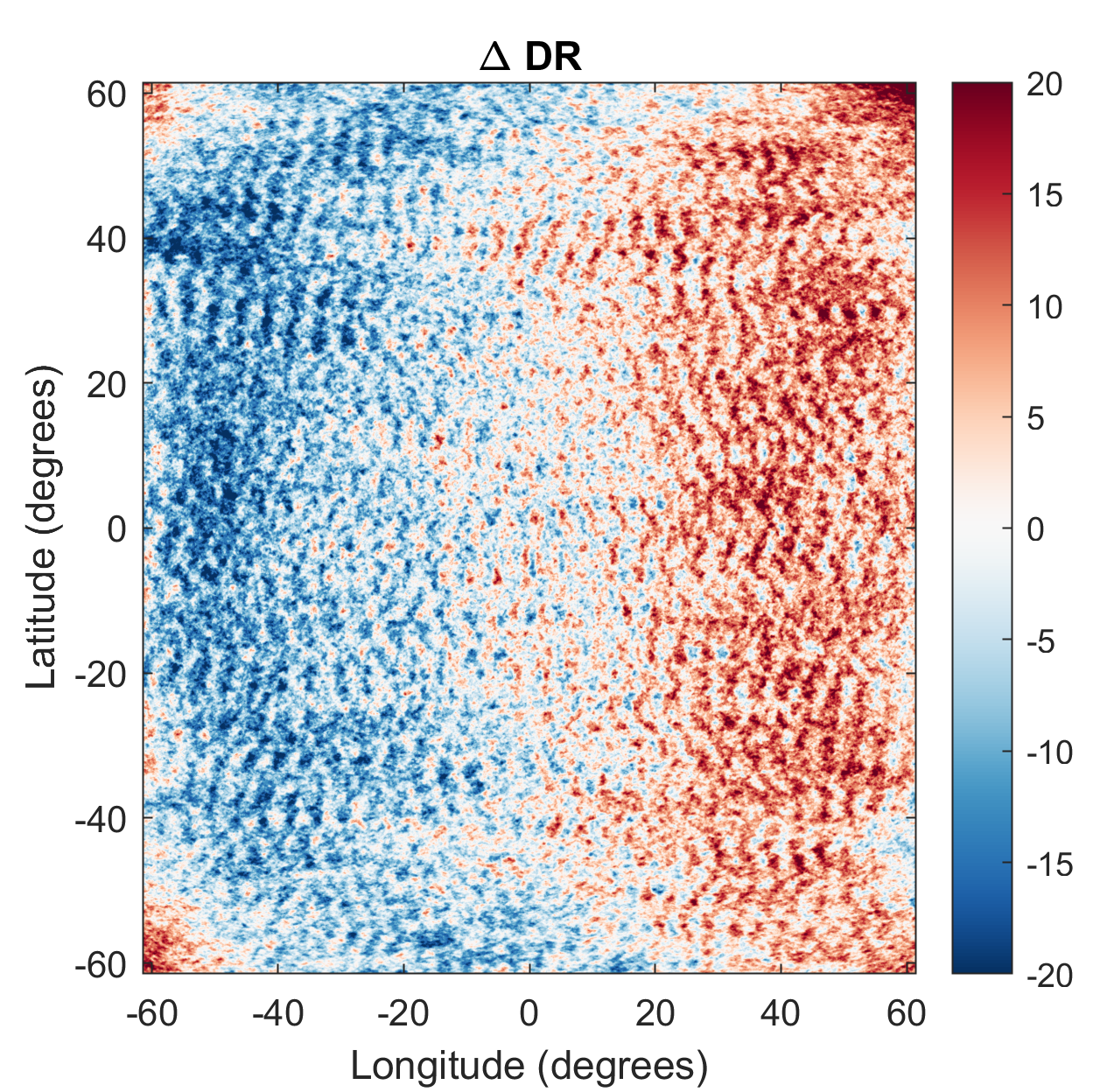}
    \caption{Rotation rate residuals (in m~s$^{-1}$) after subtracting the average rotation rate at each latitude clearly show East-West asymmetry due to the so called center-to-limb effect.}
    \label{fig:ctol_dr}
\end{figure}

In order to characterize the center-to-limb variation, we analyze the flow maps at very low B-angles $|$CRLT\_OBS$|<0.25$ degrees. The rotation rate residuals after subtraction of the average rotation rate at each latitude show a dependence on longitude (see Fig.\ \ref{fig:ctol_dr}) that is characteristic of a center-to-limb effect. This East-West asymmetry in rotation rate doesn't seem to diminish at high latitudes and hence it is a bit different from what a true center-to-limb variation would look like. As in several other studies of flow measurements before \citep{2012ApJ...749L...5Z,2017ApJ...849..144C,2021ApJ...917..100M} a similar variation seems to contaminate the measurement of meridional flow (making it peak at higher latitude with a higher amplitude). 

\begin{figure}[ht]
    \centering
    \includegraphics[width = 15 cm]{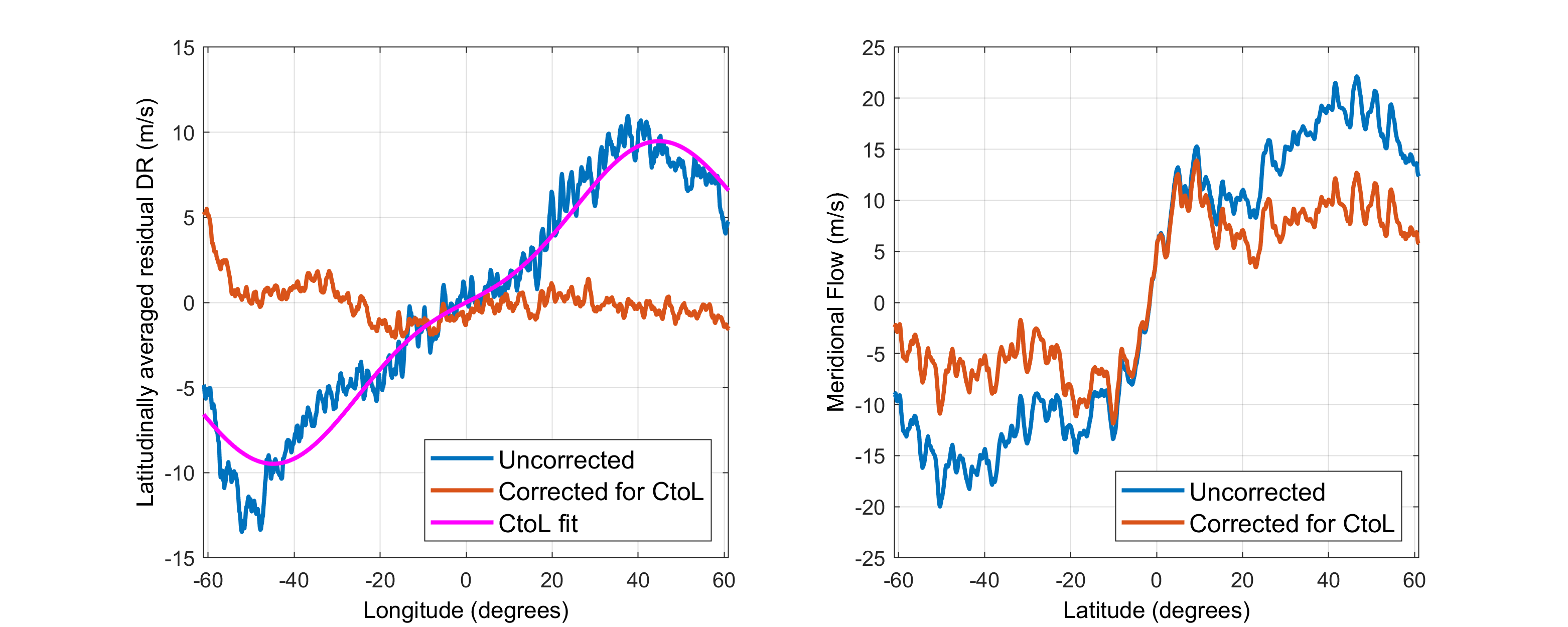}
    \caption{\textit{Left:} The residual rotation rate within $20\degr$ of the equator (blue) is fit with associated Legendre polynomials (magenta) to measure the center-to-limb effect. The corrected residual rotation rate after subtracting the center-to-limb effect is shown in red. \textit{Right:} The polynomial fit for East-West asymmetric center-to-limb effect is subtracted from meridional flow (blue) to correct it (red). }
    \label{fig:ctol_correction}
\end{figure}

\citet{2021ApJ...917..100M} found this variation to be nearly identical in both latitudinal and longitudinal directions. We, therefore, choose to characterize this asymmetry in the longitudinal direction within $20\degr$ latitude from the equator by fitting Associated Legendre Polynomials of order $m=1$ and degrees $l=1$ to $6$. This polynomial fit to the East-West asymmetry in rotation rate is also applied as a correction in the latitudinal direction to the meridional flow velocities as well \citep{2017ApJ...849..144C,2021ApJ...917..100M}. The rotation rate at low latitudes as well as the meridional flow are shown in Fig.\ \ref{fig:ctol_correction} both pre- and post-correction.

This center-to-limb effect is defined with respect to the center of the solar disk and as the Carrington latitude at the center of the solar disk changes throughout the year, the correction maps need to be interpolated to match the allignment of the solar disk.

\subsection{Other Systematic Corrections}
A P-angle correction determined during the transit of Venus in 2012 was applied to data prior to 2012, July 18 \citep{2016SoPh..291.1887C}, because the time--distance helioseismology pipeline results were not retrospectively re-processed after the determination of this correction. Annual variations in the velocity measurements were removed by fitting sine functions to the data. In order to calculate mean values of flows unbiased by the asymmetric sampling of values in the tails of their statistical distributions, we removed outliers $>2\sigma$ away and iteratively converged to the sample mean.

\end{CJK*}
\end{document}